\RequirePackage{lineno}
\documentclass[prd,twocolumn,amsmath,amssymb]{revtex4}
\usepackage{graphicx}
\usepackage{dcolumn}
\usepackage{bm}
\usepackage{epsfig}
\usepackage{overpic}
\usepackage{verbatim}
\usepackage{xspace}
\newsavebox{\tablebox}
\usepackage[colorlinks,linkcolor=blue,anchorcolor=red,citecolor=blue]{hyperref}
\usepackage{rotating}
\graphicspath{{figures/}}

\newcommand{\Rmnum}[1]{\uppercase\expandafter{\romannumeral #1}}
\newcommand{\BR}{{\cal B}}

\newcommand{\EE}{e^+e^-}
\newcommand{\MM}{\mu^+\mu^-}
\def\gevcc{\ifmmode {\mathrm{Ge\kern -0.1em V}/c^2}\else
                   {\textrm{Ge\kern -0.1em V}/$c^2$}\fi}%
\def\gev{\ifmmode {\mathrm{Ge\kern -0.1em V}}\else
                   {\textrm{Ge\kern -0.1em V}}\fi}%
\def\mevcc{\ifmmode {\mathrm{Me\kern -0.1em V}/c^2}\else
                   {\textrm{Me\kern -0.1em V}/$c^2$}\fi}%
\def\mev{\ifmmode {\mathrm{Me\kern -0.1em V}}\else
                   {\textrm{Me\kern -0.1em V}}\fi}%

\newcommand{\bfg}{\begin{figure}}
\newcommand{\efg}{\end{figure}}
\newcommand{\bitm}{\begin{itemize}}
\newcommand{\eitm}{\end{itemize}}
\newcommand{\bnum}{\begin{enumerate}}
\newcommand{\enum}{\end{enumerate}}
\newcommand{\btbl}{\begin{table}}
\newcommand{\etbl}{\end{table}}
\newcommand{\btbu}{\begin{tabular}}
\newcommand{\etbu}{\end{tabular}}
\newcommand{\beq}{\begin{equation}}
\newcommand{\edq}{\end{equation}}

\def\cpf{F_{+}}
\def\cpfpi{F_{+}^{4\pi}}
\def\fpi{\pi^{+}\pi^{-}\pi^{+}\pi^{-}}
\def\Fpi{4\pi}
\def\Nst{N_{\textrm{ST}}}
\def\Ndt{N_{\textrm{DT}}}
\def\dE{\Delta E}
\def\mbc{m_{\textrm{BC}}}

\def\missm{M^2_{\textrm{miss}}}

\newcommand{\ks}{K_S^0}
\newcommand{\kl}{K_L^0}
\newcommand{\ksl}{K_{S,L}^0}
\newcommand{\pipi}{\pi^{+}\pi^{-}}

\newcommand{\Dz}{D^{0}}
\newcommand{\Dzbar}{\bar{D}^{0}}

\newcommand{\psipp}{\psi(3770)}
\newcommand{\jpsi}{J/\psi}
\newcommand{\invfb}{\textrm{fb}^{-1}}
\def\piz{\pi^0}

\begin{document}
\normalsize
\parskip=5pt plus 1pt minus 1pt

\title{\boldmath Measurement of the $C\!P$-even fraction of $\Dz\to\fpi$}
\author{\small
M.~Ablikim$^{1}$, M.~N.~Achasov$^{11,b}$, P.~Adlarson$^{70}$, M.~Albrecht$^{4}$, R.~Aliberti$^{31}$, A.~Amoroso$^{69A,69C}$, M.~R.~An$^{35}$, Q.~An$^{66,53}$, X.~H.~Bai$^{61}$, Y.~Bai$^{52}$
% ...
\\
\vspace{0.2cm}
(BESIII Collaboration)\\
\vspace{0.2cm} {\it
$^{1}$ Institute of High Energy Physics, Beijing 100049, People's Republic of China\\
% ...
$^{4}$ Bochum Ruhr-University, D-44780 Bochum, Germany\\
% ...
$^{11}$ G.I. Budker Institute of Nuclear Physics SB RAS (BINP), Novosibirsk 630090, Russia\\
% ...
$^{31}$ Johannes Gutenberg University of Mainz, Johann-Joachim-Becher-Weg 45, D-55099 Mainz, Germany\\
% ...
$^{35}$ Liaoning Normal University, Dalian 116029, People's Republic of China\\
% ...
$^{52}$ Southeast University, Nanjing 211100, People's Republic of China\\
$^{53}$ State Key Laboratory of Particle Detection and Electronics, Beijing 100049, Hefei 230026, People's Republic of China\\
% ...
$^{61}$ University of Jinan, Jinan 250022, People's Republic of China\\
% ...
$^{66}$ University of Science and Technology of China, Hefei 230026, People's Republic of China\\
% ...
$^{69}$ University of Turin and INFN, (A)University of Turin, I-10125, Turin, Italy; (B)University of Eastern Piedmont, I-15121, Alessandria, Italy; (C)INFN, I-10125, Turin, Italy\\
$^{70}$ Uppsala University, Box 516, SE-75120 Uppsala, Sweden\\
% ...
$^{b}$ Also at the Novosibirsk State University, Novosibirsk, 630090, Russia\\
% ...
}
}

\begin{abstract}

A measurement of the $C\!P$-even fraction of the decay $\Dz\to\fpi$ is performed with a quantum-correlated $\psipp\to D\bar{D}$ data sample collected by the BESIII experiment, corresponding to an integrated luminosity of 2.93 $\invfb$.
Using a combination of $C\!P$ eigenstates, $D \to \pipi\piz$ and $D \to \ksl\pipi$ as tagging modes, the $C\!P$-even fraction is measured to be $\cpfpi = 0.735 \pm 0.015 \pm 0.005$, where the first uncertainty is statistical and the second is systematic. This is the most precise determination of this quantity to date. It provides valuable model-independent input for the measurement of the CKM angle $\gamma$ with $B^\pm\to D K^\pm$ decays, and for time-dependent studies of $C\!P$ violation and mixing in the $\Dz$-$\Dzbar$ system. 

\end{abstract}


\maketitle

\section{INTRODUCTION}

Studies of $C\!P$ violation in the heavy-quark sector can improve our understanding of the weak interaction and probe for new effects that cannot be accommodated within the Standard Model of particle physics. Many of these studies use neutral charm mesons, with a particular case of interest being when the charm meson decays to a $C\!P$ eigenstate. 
Observables in the process 
$B^\pm \to D K^\pm $,  where the $D$ is a superposition of $\Dz$ and $\Dzbar$ and is here reconstructed in a $C\!P$ eigenstate, are sensitive to the CKM angle $\gamma$, and thus allow for tests of the Standard Model description of $C\!P$ violation~\cite{GRONAU1991483,Gronau:1991dp}.  Furthermore, measurements of the decay time of charm mesons produced in a known flavor eigenstate and reconstructed in a $C\!P$ eigenstate bring information on $\Dz$-$\Dzbar$ 
 oscillations~\cite{Lenz:2020awd}. These strategies can be extended to so-called quasi $C\!P$ eigenstates: self-conjugate multi-body decays  that are not themselves $C\!P$ eigenstates, but dominated by intermediate states of definite $C\!P$~\cite{Nayak:2014tea,Malde:2015xra}.  The parameter $F_+$ quantifies the fractional $C\!P$-even content of these modes, with $F_+ = 1$ ($0$) corresponding to a pure $C\!P$-even (-odd) final state. Measurements using these quasi $C\!P$ eigenstates require good knowledge of $F_+$ for interpreting the results. 

$C\!P$-even fractions and other parameters associated with the strong dynamics of charm-meson decays are preferentially measured from data collected at the $\psipp$ resonance. The quantum correlation between the two charm mesons produced from the decay of the $\psipp$ gives rise to interference effects that provide access to these parameters.  This approach has the benefit of being model-independent and is to be contrasted with alternative strategies that are based on amplitude models constructed from decays of mesons produced in incoherent environments~\cite{BaBar:2010uep,Belle:2010xyn}.  Measurements of the $C\!P$-even fraction with correlated $D\bar{D}$ pairs have been performed with data collected by the CLEO-c experiment for 
a range of decay modes including $D \to \fpi$
~\cite{Nayak:2014tea,Malde:2015mha,Harnew:2017tlp,Resmi:2017fuo}.

The decay $D \to \fpi$ is an attractive mode for $C\!P$-violation studies on account of its reasonably high branching fraction and the good reconstruction efficiency that both LHCb and Belle~II have for the fully charged final state~\cite{LHCb:2016gpc,LHCb:2019yan}.
This paper presents a determination of the $C\!P$-even fraction, $\cpfpi$, of this decay using quantum-correlated $\psipp \to D\bar{D}$ data collected by the BESIII experiment, corresponding to an integrated luminosity of 2.93~$\invfb$. It is the first measurement of a $C\!P$-even fraction with this data set, although BESIII has presented studies of other hadronic parameters performed with correlated $D\bar{D}$ pairs in a variety of channels~\cite{BESIII:2014rtm,BESIII:2020hlg,BESIII:2020lpk,KsKK,Ablikim:2021cqw}.  

\section{MEASUREMENT METHOD}
\label{sec:method}

The strong decay $\psipp \to D\bar{D}$  conserves the negative C quantum number, leaving the charm-meson pair in an overall $P$-wave state. The quantum correlation between the two mesons in this antisymmetric wavefunction  provides the opportunity to determine the $C\!P$-even fraction of $\Dz\to\fpi$.
(The parameter for the $\Dzbar$ decay is identical.)

The method adopted in this analysis is similar to that applied in previous measurements~\cite{Nayak:2014tea,Malde:2015mha,Resmi:2017fuo},
and is based on samples of  `double-tag' (DT)  and `single-tag' (ST) events.
The DT events are those where one $D$ meson decays to the signal final-state $\fpi$ and the $\bar{D}$ meson decays to a `tag mode'.
The ST events are those where only one charm meson is reconstructed in its decay to a tag mode.

Various classes of tag modes are employed, the first ones being pure $C\!P$ eigenstates. Several decays involving $\ks$ and $\kl$ mesons are included in this category, as $C\!P$ violation in the kaon system can be neglected at the current level of experimental precision. The same consideration applies to $C\!P$ violation associated with the charm mesons themselves~\cite{Amhis:2019ckw}. For a $C\!P$-tag mode, here denoted as $f$, the predicted DT yield is given by
\beq
\label{eq:cpDT}
\begin{split}
&M(\Fpi,f) = \\ 
2 & N_{D\bar{D}}\BR(\Fpi)\BR(f)\epsilon_{\textrm{DT}}[1-\eta_{C\!P}^f(2\cpfpi-1)]\, ,
\end{split}
\edq
where $N_{D\bar{D}}$ is the number of neutral charm-meson pairs in the sample, $\BR(X)$ is the branching fraction of $D \to X$, $\epsilon_{DT}$ is the DT reconstruction efficiency of the event, and $\eta_{CP}^f$ is the $C\!P$-eigenvalue ($+1$ or $-1$) of the tag channel $f$.  In this and subsequent expressions, terms of ${\cal O}(y^2)$ in the charm mixing parameter $y = \left(0.615^{+0.056}_{-0.055} \right)\%$~\cite{Amhis:2019ckw} are neglected.

The predicted ST yield is given by 
\beq
\label{eq:cpST}
S(f) = 2 N_{D\bar{D}}\BR(f)\epsilon_{\textrm{ST}}[1-\eta_{C\!P}^fy],
\edq
where $\epsilon_{ST}$ is the reconstruction efficiency of the ST decay.
Then $\cpfpi$ can be accessed through
\beq
\label{eq:cpDTST}
\cpfpi = \frac{N^+}{N^+ + N^-}\, , 
\edq
where $N^+$, which measures the proportion of $C\!P$-even $D$ mesons that decay to $\fpi$, is 
$M(\Fpi,f)[1-\eta_{C\!P}^f y]  / S(f)$ for a $C\!P$-odd tag, 
and $N^-$ is the analogous quantity for a $C\!P$-even tag. Experimentally, $N^+$ and $N^-$ are determined from the corresponding ratios involving the  measured DT and ST yields and are averaged over all employed tags.  This calculation assumes that the DT-event efficiency can be factorized into the ST efficiency and the reconstruction efficiency of the signal decay and that any  corrections to this assumption are common to all classes of DT.  Possible exceptions to these assumptions are examined in the assignment of systematic uncertainties. Modifications to this procedure, described in Sec.~\ref{sec:evtsel}, are required for certain classes of tag involving $\kl$ mesons where the ST yields cannot be directly measured.

A specific tag used in the analysis is $D \to \pipi\piz$, which is a quasi $C\!P$ eigenstate.
Here the predicted DT yield is given by
\beq
\label{eq:pipipizDT}
\begin{split}
&M(\Fpi,f) = \\ 
& 2 N_{D\bar{D}}\BR(\Fpi)\BR(\pi\pi\piz)\epsilon_{\textrm{DT}}[1-(2\cpf^{\pi\pi\piz}-1)(2\cpfpi-1)]\, ,
\end{split}
\edq
where $\cpf^{\pi\pi\pi^0} = 0.973 \pm 0.017$ is the $C\!P$-even fraction of the tag mode~\cite{Malde:2015mha},
and the predicted ST yield is 
\beq
\label{eq:pipipizST} 
S(\pi\pi\piz) = 2 N_{D\bar{D}}\BR(\pi\pi\piz)\epsilon_{\textrm{ST}}[1-(2\cpf^{\pi\pi\piz}-1)y]\, .
\edq
In this case,  $\cpfpi$ can be accessed through the ratio
\beq
\label{eq:pipipizDTST}
\cpfpi = \frac{N^+\cpf^{\pi\pi\piz}}{N^{\pi\pi\piz} - N^+ + 2N^+\cpf^{\pi\pi\piz}},
\edq
with $N^{\pi\pi\piz}=M(\Fpi,\pi\pi\piz)[1-(2\cpf^{\pi\pi\piz}-1)y]/{S(\pi\pi\piz)}$. Experimentally, $N^+$ and $N^{\pi\pi\piz}$ are again determined from the measured DT and ST yields, with the same assumptions concerning the factorization of the DT-event efficiency  as discussed above.

The self-conjugate decays $D \to \ks\pipi$ and $D \to \kl\pipi$ are known to have $C\!P$-even fractions close to 0.5~\cite{Gershon:2015xra}, and hence would give poor sensitivity to $\cpfpi$ if used as a global tag in an analogous manner to $D \to \pipi\piz$. However, a local measurement may be performed where the position of the tag decay in its phase space is considered.  This is possible because studies on the variation in the strong-phase difference between the $\Dz$ and $\Dzbar$ amplitudes across the Dalitz plot have been performed for these modes.  In particular, measurements exist of the amplitude-weighted cosine of the average strong-phase difference in binned regions of phase space~\cite{BESIII:2020lpk,BESIII:2020hlg,CLEO:2010iul}. 
The predicted DT yields in bin $i$ are given by 
\begin{eqnarray}
M_i(\Fpi,\ks\pipi) = \hspace*{3.0cm} & \nonumber \\
H[K_i+K_{-i} - 2\sqrt{K_iK_{-i}}c_i(2\cpfpi-1)]  \, , \hspace*{0.2cm} &  \nonumber \\
 M^\prime_i(\Fpi,\kl\pipi) = \hspace*{3.0cm} & \nonumber\\
 H^\prime [K^\prime_i+K^\prime_{-i} + 2\sqrt{K^\prime_iK^\prime_{-i}}c^\prime_i(2\cpfpi-1)]  \, , \hspace*{0.1cm} &
\label{eq:KsKlDT}
\end{eqnarray}
where $H$ and $H^\prime$ are  normalization factors, $K_i$ ($K^\prime_i$) is the fractional rate of $\Dz$ decays to $\ks\pipi$ ($\kl\pipi$) in the $i$-th bin,
and $c_i$ ($c^\prime_i$) is the amplitude-weighted cosine of the average strong-phase difference in the ~ $i$-th bin.  When comparing the observed distribution of events throughout the bins with the predictions, there is no need to know the overall efficiency scale. However, it is important to account for relative variations in DT efficiency bin-to-bin as well as migration effects brought about by bin misassignments. 

\section{BES\Rmnum{3} DETECTOR AND MONTE CARLO SIMULATION}
\label{detmc}

The BESIII detector~\cite{Ablikim:2009aa} records symmetric $e^+e^-$ collisions 
provided by the BEPCII storage ring~\cite{Yu:2016cof}, which operates with a peak luminosity of $1\times10^{33}$~cm$^{-2}$s$^{-1}$
in the center-of-mass energy range from 2.0 to 4.95~$\gev$.
BESIII has collected large data samples in this energy region~\cite{BESIII:2020nme}. The cylindrical core of the BESIII detector covers 93\% of the full solid angle and consists of a helium-based
 multilayer drift chamber~(MDC), a plastic scintillator time-of-flight
system~(TOF), and a CsI(Tl) electromagnetic calorimeter~(EMC),
which are all enclosed in a superconducting solenoidal magnet
providing a 1.0~T magnetic field. The solenoid is supported by an
octagonal flux-return yoke with resistive plate counter muon-identification modules interleaved with steel. 
The charged-particle momentum resolution at $1~{\rm GeV}/c$ is
$0.5\%$, and the 
${\rm d}E/{\rm d}x$
resolution is $6\%$ for electrons
from Bhabha scattering. The EMC measures photon energies with a
resolution of $2.5\%$ ($5\%$) at $1$~GeV in the barrel (end-cap)
region. The time resolution in the TOF barrel region is 68~ps, while
that in the end-cap region is 110~ps.

Simulated samples, which are produced with the {\sc geant4}-based~\cite{geant4} Monte Carlo (MC) package that includes the description of the detector geometry and response,
are used to determine the detection efficiencies and estimate the backgrounds.
The beam-energy spread of 0.97~MeV and the initial-state radiation (ISR) in the $\EE$ annihilations, which is modeled with the generator {\sc kkmc}~\cite{kkmc}, are included in the simulation.
The inclusive MC samples for background studies consist of the production of neutral and charged charm-meson pairs from $\psipp$ decays, decays of the $\psipp$ to charmonia or light hadrons, the ISR production of the $\jpsi$ and $\psipp$ states, continuum processes, and the QED processes $\EE\to\EE$, $\EE\to\MM$ and $\EE\to\tau^+\tau^-$.  No attempt is made to include quantum-correlation effects in $\psipp \to D\bar{D}$ decays in the inclusive MC sample.
The equivalent integrated luminosity of the inclusive MC samples is about 10 times that of the data, apart from the production of $D\bar{D}$ events, where the equivalent integrated luminosity is about 20 times that of the data.
All particle decays are modeled with {\sc
evtgen}~\cite{evtgen} using branching fractions either taken from the
Particle Data Group~\cite{pdg}, when available,
or otherwise estimated with {\sc lundcharm}~\cite{lundcharm,Yang:2014vra}.
The final-state radiation from the charged final-state particles is incorporated with the {\sc photos} package~\cite{photos}.

Signal MC samples of around 200,000 events are generated separately for the different tag channels. 
In this generation, the $D\to\fpi$ decay follows an isobar-based amplitude model  fitted to a BESIII sample of these decays, where the flavor of the decaying meson is inferred by reconstructing the other charm meson in the event through its decay into a flavor-specific final state. The model contains the main resonant structures observed in data.  The simulated DT samples involving $D \to \ksl\pipi$ tags are an order of magnitude larger in size and the tag decays are implemented with an amplitude model developed by the BaBar collaboration~\cite{BaBar:2010nhz}. In all the DT samples, quantum correlations are included in the generation to ensure the best possible description of the reconstruction efficiency, especially for the different bins of phase space of the $\ksl\pipi$ tag modes.

\section{EVENT SELECTION AND YIELD DETERMINATION}
\label{sec:evtsel}

Table~\ref{tab:TagInfo} lists the tag categories that are employed in the analysis: $C\!P$-even eigenstates, $C\!P$-odd eigenstates, the quasi-$C\!P$-even mode $D\to \pipi\piz$, and the self-conjugate decays $D\to \ksl\pipi$ of mixed $C\!P$ that are analyzed in bins of phase space.
The final-state resonances are reconstructed in the following channels:
$\ks\to\pipi$, $\piz\to\gamma\gamma$, $\eta\to\gamma\gamma$ and $\pipi\piz$, $\omega\to\pipi\piz$, and  $\eta'\to\pipi\eta$ and $\gamma\pipi$.

\begin{table}[!hbp]
\renewcommand\arraystretch{1.5}
\centering
\caption{Summary of tag channels.\vspace*{0.2cm}}
\label{tab:TagInfo}
\begin{tabular}{ l  c }
\toprule
Category & Decay modes\\
\colrule
	$C\!P$-even    &  $K^{+}K^{-}$, $\ks\piz\piz$, $\kl\piz$, $\kl\omega$ \\
	$C\!P$-odd     &  $\ks\piz$, $\ks\eta$, $\ks\eta'$, $\ks\omega$, $\kl\piz\piz$ \\
	Quasi $C\!P$-even & $\pipi\piz$  \\
	Mixed $C\!P$ & $\ks\pipi$, $\kl\pipi$ \\
\botrule
\end{tabular}
\end{table}

DT events formed of tags involving a $\kl$ are partially reconstructed using a missing-mass technique. All other classes of DT events are fully reconstructed, as are ST events that do not contain a $\kl$ meson.

\subsection{Basic event selection}

Several basic requirements are imposed to ensure the quality of charged tracks in the analysis.  Those tracks from $\ks$ candidates must lie within 20~cm of $\EE$ interaction point (IP) along the $z$-axis, 
which is the symmetry axis of the MDC. 
All other tracks must have a point of closest approach to the IP within $\pm$10~cm along the $z$-axis and 1~cm in the transverse plane.
The polar angle $\theta$ with respect to the axis of the drift chamber must satisfy the condition $|\textrm{cos}\theta|<0.93$.
Charged pions and kaons are distinguished by combining the information of the flight time measured from the TOF and the d$E$/d$x$ measured in the MDC. 
The corresponding probabilities $P_K$ ($P_\pi$) for the $K$ ($\pi$) hypothesis are calculated and 
the track is labeled as a $K$ ($\pi$) candidate if $P_K > P_\pi$ ($P_K < P_\pi$).

The photon candidates are reconstructed from the showers in EMC, with the energy required to be larger than 25~MeV for  
barrel showers ($|\!\cos\theta|<0.80$) and 50~MeV for end-cap showers ($0.86<|\!\cos\theta|<0.92$).
To suppress electronic noise or activity unrelated to the events, the time of the cluster measured from the EMC is required to be within 0 and 700~ns after the event start time.
Furthermore, to eliminate showers originating from charged tracks, the angle subtended by the EMC shower and the position of the closest charged track at the EMC must be greater than 20 degrees as measured from the IP.

The $\ks$ candidates are reconstructed from pairs of tracks with opposite charge on which no particle-identification requirements are imposed.
A fit is applied to constrain the track pair to a common vertex.
The flight significance $L/\sigma_{L}$ is required to be larger than 2, 
where $L$ is the flight distance and $\sigma_{L}$ is the corresponding standard deviation.
In addition, the invariant mass is required to be within [0.487,0.511]~$\gevcc$.

Pairs of photons are used to reconstruct $\piz(\eta)$ candidates, where the invariant mass of the pair must lie within [0.115, 0.150] ([0.480, 0.580])~$\gevcc$ and at least one photon candidate is from the barrel region.
In order to improve the momentum resolution, a kinematic fit is performed, where the reconstructed $\piz(\eta)$ invariant mass is constrained to the known value~\cite{pdg}, and the fitted momentum of the $\piz(\eta)$ is used in the subsequent stages of the analysis. 
When reconstructing $\eta\to\pipi\piz$ decays, the invariant mass of the $\eta$ candidate is required to be within [0.530, 0.565] $\gevcc$. 
Similarly, when reconstructing $\omega\to\pipi\piz$, $\eta'\to\pipi\eta(\gamma\gamma)$ and $\eta'\to\gamma\pipi$, the invariant mass is required to lie within [0.750,0.820], [0.940, 0.976] and [0.940, 0.970] $\gevcc$, respectively.

\begin{figure*}[!htp]
\begin{center}
\begin{overpic}[width=1.0\textwidth]{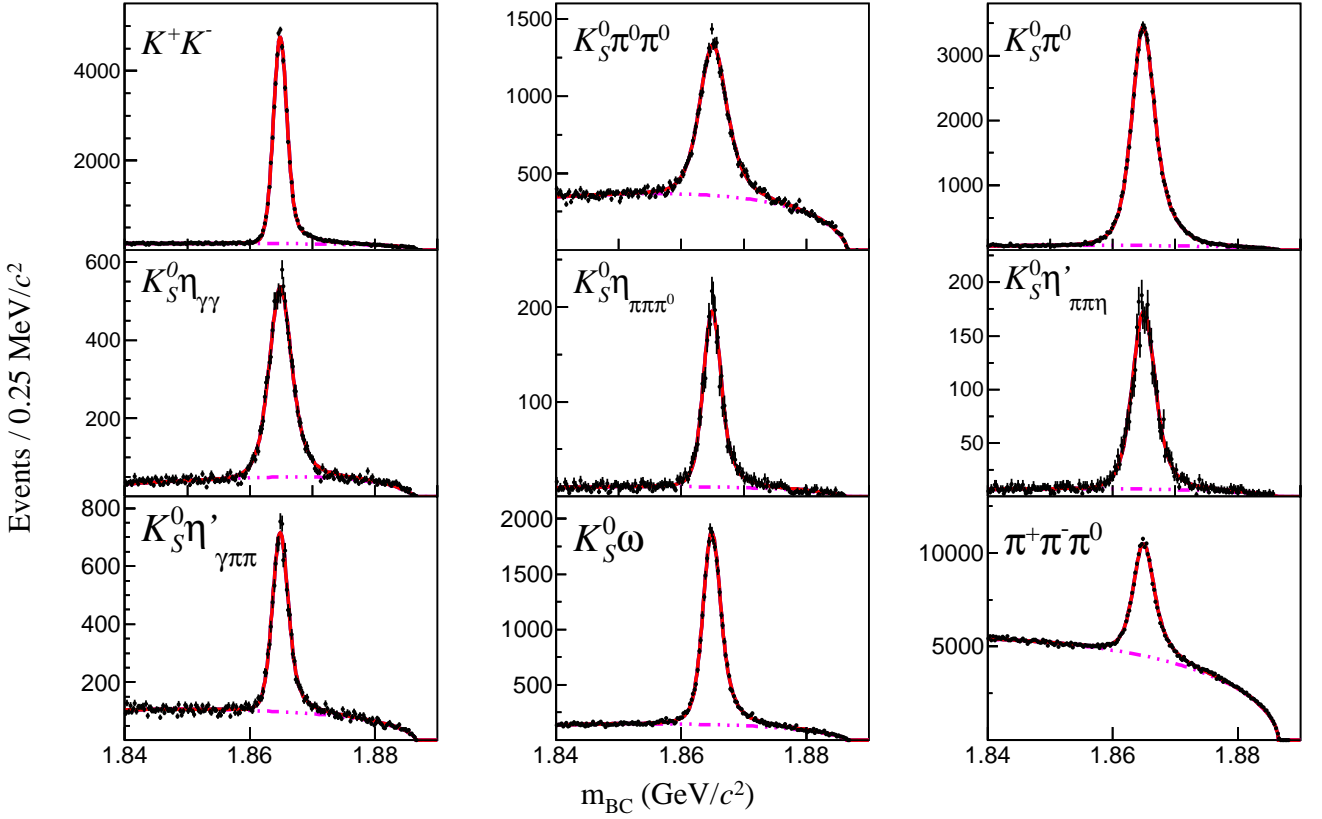}\end{overpic}
\caption[]{Distributions of $\mbc$ and fits used to determine the ST yields. 
In each plot the black dots with error bar are data, 
the total fit result is shown as the red solid line and  
the continuum background is shown as the pink dashed line.
}
\label{fig:STFit}
\end{center}
\end{figure*}

\subsection{ST event selection and yields}

ST events that do not contain a $\kl$ meson are reconstructed from the charged pion, charged kaon, $\ks$, and resonance candidates. 
To suppress the combinatorial background, the energy difference, $\dE = E_D -\sqrt{s}/2$, is required to be within $\pm3\sigma_{\dE}$ of the $\dE$ peak, 
where $E_D$ is the measured energy of the $D$-meson candidate in the center-of-mass frame, $\sqrt{s}$ is the center-of-mass energy, and $\sigma_{\dE}$ is the resolution of the $\dE$ distribution.
To suppress background from cosmic and Bhabha events in  the tag channel $D\to K^+K^-$, the two charged tracks must have a TOF time difference of less than 5~ns, and the further requirement that neither track is identified as an electron or a muon is applied.
To suppress $D \to \ks\piz$ contamination in the selection of $D \to \pipi\piz$ decays,
a $\ks$-veto is applied, in which the event is rejected 
if the charged pion pair has a significant flight distance ($L/\sigma_{L}>2$) and invariant mass lying within the range [0.481, 0.514]$\gevcc$. This use of flight-distance information is preferred to removing all events within the $\ks$ mass window as the $C\!P$-even fraction of this mode is only known for the full phase space of the decay~\cite{Malde:2015mha}.

In order to determine the ST yields for the $C\!P$ eigenstate and quasi-$C\!P$ eigenstate tags, the beam-constrained mass, $\mbc = \sqrt{(\sqrt{s}/2)^2 - |\bf{p_D}|^2}$, is used to identify the signal, where $\bf{p_D}$ is the three-momentum vector of the $D$ candidate in the center-of-mass frame. 
Binned maximum-likelihood fits are performed on the $\mbc$ distributions, as shown in Fig.~\ref{fig:STFit}, 
where the signal is described by the shape found in MC simulation convolved with a Gaussian function, to account for differences in resolution between MC and data, and the combinatorial background is described by an ARGUS function with an end-point fixed to $\sqrt{s}/2$~\cite{ARGUS:1990hfq}.
The peaking-background contributions from other charm decays are estimated from MC simulation and then subtracted from the fitted ST yields. In the case of  $D \to \ks\omega$, $\ks\eta_{\pi\pi\piz}$, and $\ks\eta'_{\gamma\pi\pi}$, the peaking-background fractions are 11.5\%, 9.7\%, and 3.9\%, respectively, and arise mainly from 
$D\to\ks\pipi\piz$ decays.
The peaking-background fraction in the $D \to \pipi\piz$ sample is 5\%, mainly from $D\to\ks\piz$ decays. The contamination in the 
$D \to \ks\piz(\piz)$ selection mainly comes from $D\to\pipi\piz(\piz)$ decays and constitutes   0.3\% (3.1\%) of the sample.   
For the other three channels, the peaking-background are not significant.
The ST yields after background subtraction, $N_{\rm ST}$, are summarized in Table~\ref{tab:STDTYields}, together with the $\dE$ window imposed for each channel.

The difficulty of reconstructing $\kl$ mesons means that it is impractical to select ST samples for the modes $D \to \kl X$ ($X=\omega$, $\piz$, or $\piz\piz$).  However, effective ST yields  can be determined through the relation
 \beq
\label{eq:KlST}
N_{\rm ST}(\kl X) = 2N_{D\bar{D}} \, \BR (\kl X) \, \epsilon'_{\,\textrm{ST}}(\kl X),
\edq
where $N_{\rm ST}(\kl X)$ is the effective ST yield for mode $D \to \kl X$, $N_{D\bar{D}} = (10597\pm28\pm98)\times 10^3$ is the number of neutral charm-meson pairs in the sample~\cite{BESIII:2018iev} and
$\BR (\kl X)$ is the branching fraction of the $D \to \kl X$ mode~\cite{BAM:567}.  The effective efficiency
$\epsilon'_{\,\textrm{ST}}(\kl X)$ is defined as the ratio of the efficiency for reconstructing $D \to \fpi$ versus $D \to \kl X$ DT events with the partial-reconstruction technique, as  discussed below, and the reconstruction efficiency for $D \to \fpi$.  The effective ST yields are included in Table~\ref{tab:STDTYields}.

\subsection{DT event selection and yields}

DT samples are selected for all tags not involving a $\kl$ by attempting to reconstruct  the $D \to \fpi$ signal decay from  the remaining tracks in the ST samples.
Only events with four extra charged tracks are considered.
All four charged tracks must be identified as pions and their net charge is required to be zero.
A tight $\ks$ veto is applied to suppress background from  $D \to \ks\pipi$ decays: 
if the invariant mass of any of the $\pipi$ pairs lies within [0.481, 0.514]$\gevcc$, the event is rejected.  As this requirement removes a specific region of phase space, rejecting around 13\% of the signal, it has the potential to bias the $C\!P$-even fraction of the sample.  This effect is considered in the assignment of systematic uncertainties.
The energy difference $\dE$ of the signal decay is required to lie within [$-0.026, 0.023$]$\gev$.

\begin{figure*}[!htp]
\begin{center}
\begin{overpic}[width=1.0\textwidth]{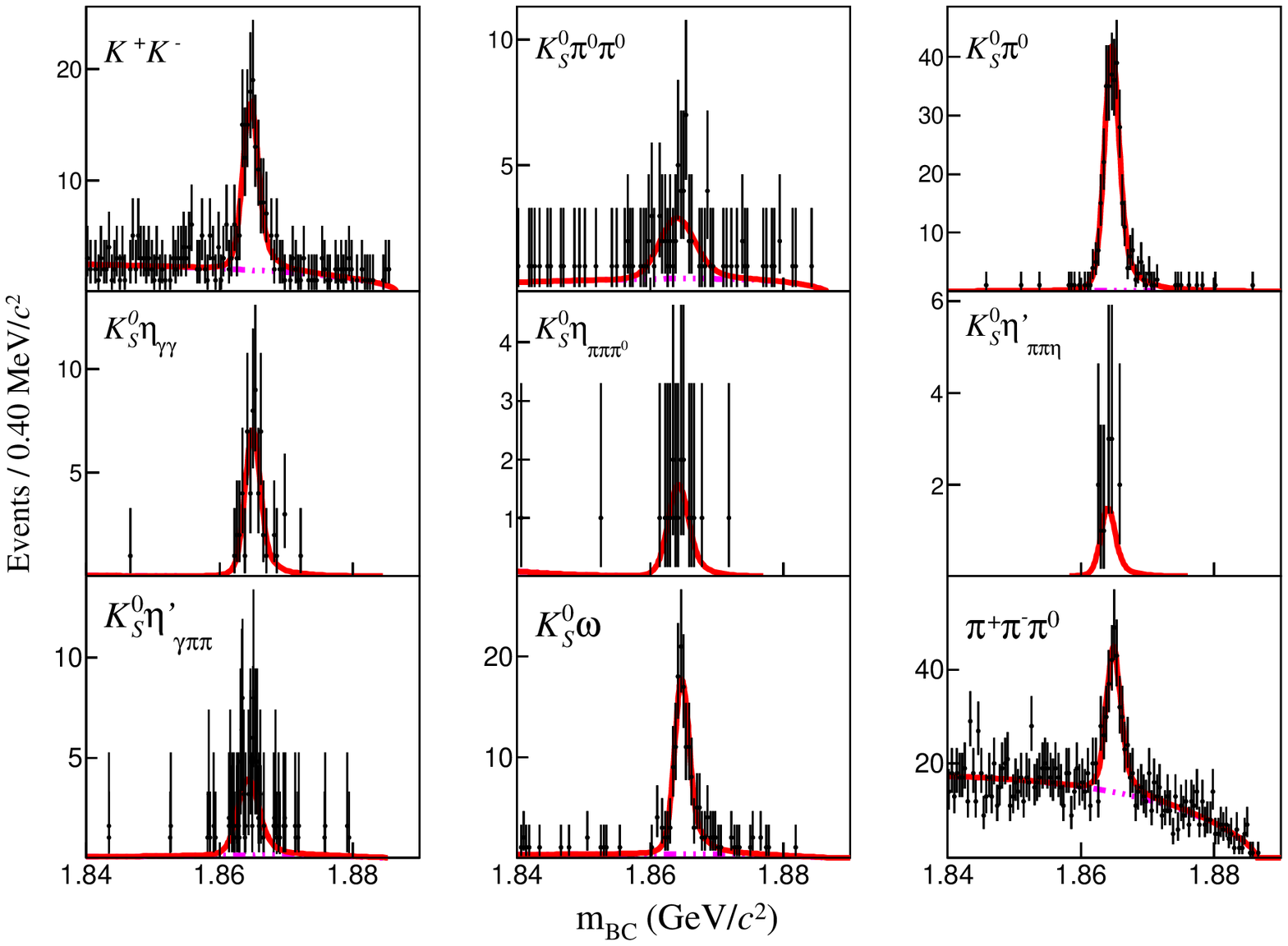}\end{overpic}
\begin{overpic}[width=1.0\textwidth]{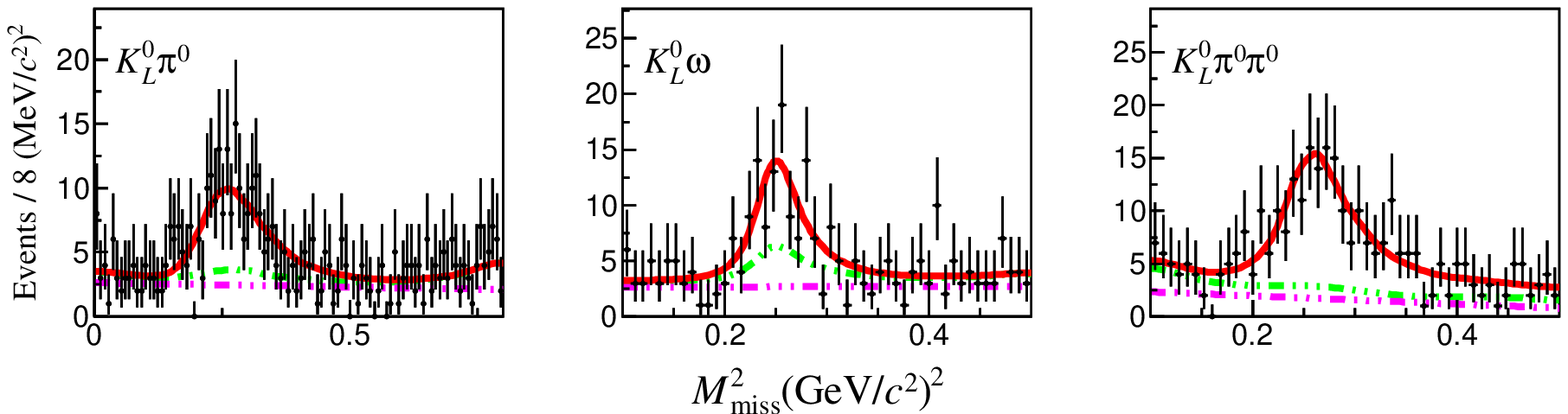}\end{overpic}
\caption[]{Distributions and fits used to determine the DT yields.  These distributions are of $\mbc$ in the fully reconstructed case, and of $\missm$ for the partially reconstructed case involving $\kl$ mesons.
In each plot the black dots with error bar are data, 
the total fit result is shown as the red solid line,
the continuum background is shown as the pink dashed line, and sum of the continuum and the peaking background for the $\kl X$ DTs is shown as the green dashed line.
}
\label{fig:DTFit}
\end{center}
\end{figure*}

For each fully reconstructed DT channel, the DT yield is determined by an unbinned maximum-likelihood fit to the $\mbc$ distribution of the signal $D$ candidate. 
The $\mbc$ distributions and the fits for the $C\!P$ eigenstate and quasi-$C\!P$ eigenstate tags are presented in Fig.~\ref{fig:DTFit}.  In these fits, the signal is described by the MC-simulated shape convolved with a Gaussian function, and the combinatorial background is described by an ARGUS function. 
The fitted signal yield includes contamination from peaking-background contributions.
The most significant source of peaking background is the residual contamination from $D \to \ks\pipi$ decays in the selection of the signal channel. 
The size of this background is determined from data by 
performing fits to the invariant-mass distribution of $\pipi$ pairs associated with the signal candidate without any $\ks$ veto in place. The results are then scaled by the suppression factor of the veto on the background mode, as determined from the MC simulation.
This background is found to vary from $2\%$ to $5\%$, depending on the tag mode.
The size of the other peaking backgrounds, such as $D \to \ks\piz$ in the $D \to \pipi\piz$ tag, are estimated from MC simulation. Where necessary, corrections are applied to account for the quantum-correlated enhancements or suppressions that are not simulated in the inclusive simulation, according to the $C\!P$ content of the background. 
The DT yields after background subtraction are summarized in Table~\ref{tab:STDTYields}.

DT channels involving a $D \to \kl X$  tag mode cannot be fully reconstructed.  Nonetheless, a partial reconstruction is performed, accounting for all other charged and neutral particles in the event, and the yield is determined using a missing-mass-squared technique.  In this procedure the $D\to\fpi$ candidate is first reconstructed with the same criteria as used previously, and then the standard selections are imposed to select the remaining charged and neutral tracks in the event. 
Candidates for the tag modes 
 $D \to \kl\piz$, $D \to \kl\piz\piz$, $D \to \kl\omega$, and $D \to \kl\pipi$ are selected from the charged tracks and $\piz$ candidates in the event.  Any events with surplus charged tracks, surplus $\piz$ candidates, or an $\eta \to \gamma\gamma$ candidate are rejected.  
 In the case of $D \to \kl\omega$, events containing more than one $\omega$ candidate are discarded.
 
 The signal yields in the partially reconstructed DTs are determined by performing unbinned maximum-likelihood fits to the squared missing-mass 
 $\missm = (\sqrt{s}/2-E_X)^2 - |{\bf p_{X}}+{\bf \hat{p}_{\Fpi}}\sqrt{s/4-M^2_D}|^2$, which is calculated in the center-of-mass frame. Here $E_X$ and ${\bf p_{X}}$ are the energy and three-momentum, respectively, of the reconstructed particles in the event not associated with the signal candidate, ${\bf \hat{p}_{\Fpi}}$ is the direction of the signal candidate, and $M_D$ is the known $\Dz$ mass~\cite{pdg}.
 The $\missm$ distributions and fits for the $C\!P$ eigenstate tags are included in  Fig.~\ref{fig:DTFit}. The resulting signal yields can be found in Table~\ref{tab:STDTYields}.  In these fits, the signal distribution is described by the shape found in the MC simulation convolved with a Gaussian function whose width is a free parameter. Fixed contributions from MC simulation are included for peaking backgrounds, and the continuum background is modeled with a first-order Chebychev polynomial, whose parameters are determined in the fit. Peaking backgrounds arise mainly from $D\to\ks\pipi$ decays that are misreconstructed as $D \to \pipi\pipi$, or decays involving an unreconstructed $\ks$ meson that contaminate the tag candidates. These  backgrounds constitute 28\%, 16\%, and 7\% of the  $D \to \kl\omega$, $\kl\piz$, and $\kl\piz\piz$ samples, respectively.

\begin{table}[!hbp]
\renewcommand\arraystretch{1.5}
\centering
\caption{Summary of $\dE$ requirements, ST yields ($\Nst$), and DT yields ($\Ndt$) for each tag mode, grouped by tag category. The uncertainties are statistical. 
The entry `/' indicates that the information is not defined ($\dE$ for $\kl X$ modes) 
or not required for the analysis (ST yields for $\ksl \pipi$).
 \vspace*{0.2cm}}
\footnotesize
\begin{tabular}{ l  c  r@{$\pm$}l r@{$\pm$}l}
\toprule
Mode  & $\dE$ (GeV) & \multicolumn{2}{c}{$\Nst$}  & \multicolumn{2}{c}{$\Ndt$}\\
\colrule
$K^{+}K^{-}$  & [$-0.021, 0.020$] & 56668&262 & 115.4&14.4\\
$\ks\piz\piz$ & [$-0.072, 0.053$] & 73176&299 & 36.4&10.3\\
$\kl\piz$ 	 & 		/					 	 & 79689&5631& 130.9&18.8\\
$\kl\omega$	 & 		/					 	 & 29128&1962& 61.5&13.8\\
\multicolumn{6}{c}{} \\
$\ks\piz$		  & [$-0.071, 0.051$] & 73176&299 & 326.0&19.2\\
$\ks\eta_{\gamma\gamma}$  & [$-0.038, 0.036$] & 10071&123 & 57.7&7.7\\
$\ks\eta_{\pi\pi\piz}$    & [$-0.035, 0.028$] & 2775&65 & 16.5&4.2\\
$\ks\eta'_{\pi\pi\eta}$   & [$-0.035, 0.031$] & 3449&67 & 11.6&3.5\\
$\ks\eta'_{\gamma\pi\pi}$ & [$-0.031, 0.025$] & 8691&126 & 41.1&7.5\\
$\ks\omega$               & [$-0.042, 0.033$] & 26220&215 & 128.7&13.8\\
$\kl\piz\piz$             & / & 25772&2184& 178.5&23.9\\
\multicolumn{6}{c}{} \\
$\pipi\piz$  & [$-0.062, 0.051$] & 115556&682 & 190.7&24.6\\
\multicolumn{6}{c}{} \\
$\ks\pipi$   & [$-0.026, 0.023$] & \multicolumn{2}{c}{/}  & 539.7&26.0\\
$\kl\pipi$ & / & \multicolumn{2}{c}{/}  & 1374.6&50.4\\
\botrule
\end{tabular}
\label{tab:STDTYields}
\end{table}

\section{$C\!P$-EVEN FRACTION MEASUREMENT}

The $C\!P$-even fraction is determined using the ST and DT yields of the $C\!P$-eigenstate and quasi-$C\!P$-eigenstate tags, and the distribution of decays in tag phase space of the $D \to \ksl \pipi$ DT events.  The analysis follows the procedure outlined in  Sec.~\ref{sec:method}.

\subsection{$C\!P$-eigenstate tags}

\begin{figure}[!htp]
\begin{center}
\begin{overpic}[width=0.46\textwidth]{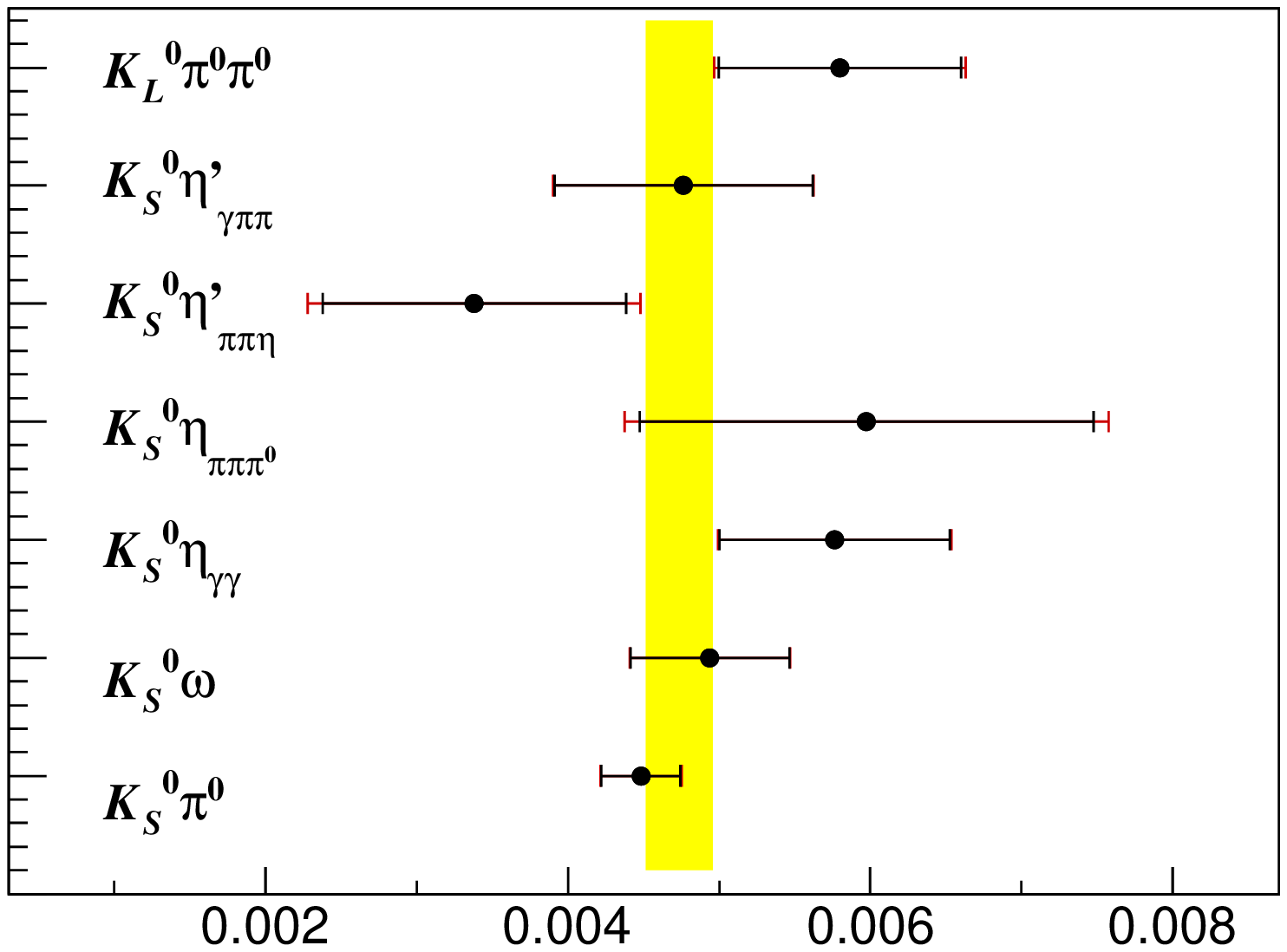}\put(75,55){(a)}\put(49,0){$N^+$}\end{overpic}
\begin{overpic}[width=0.46\textwidth]{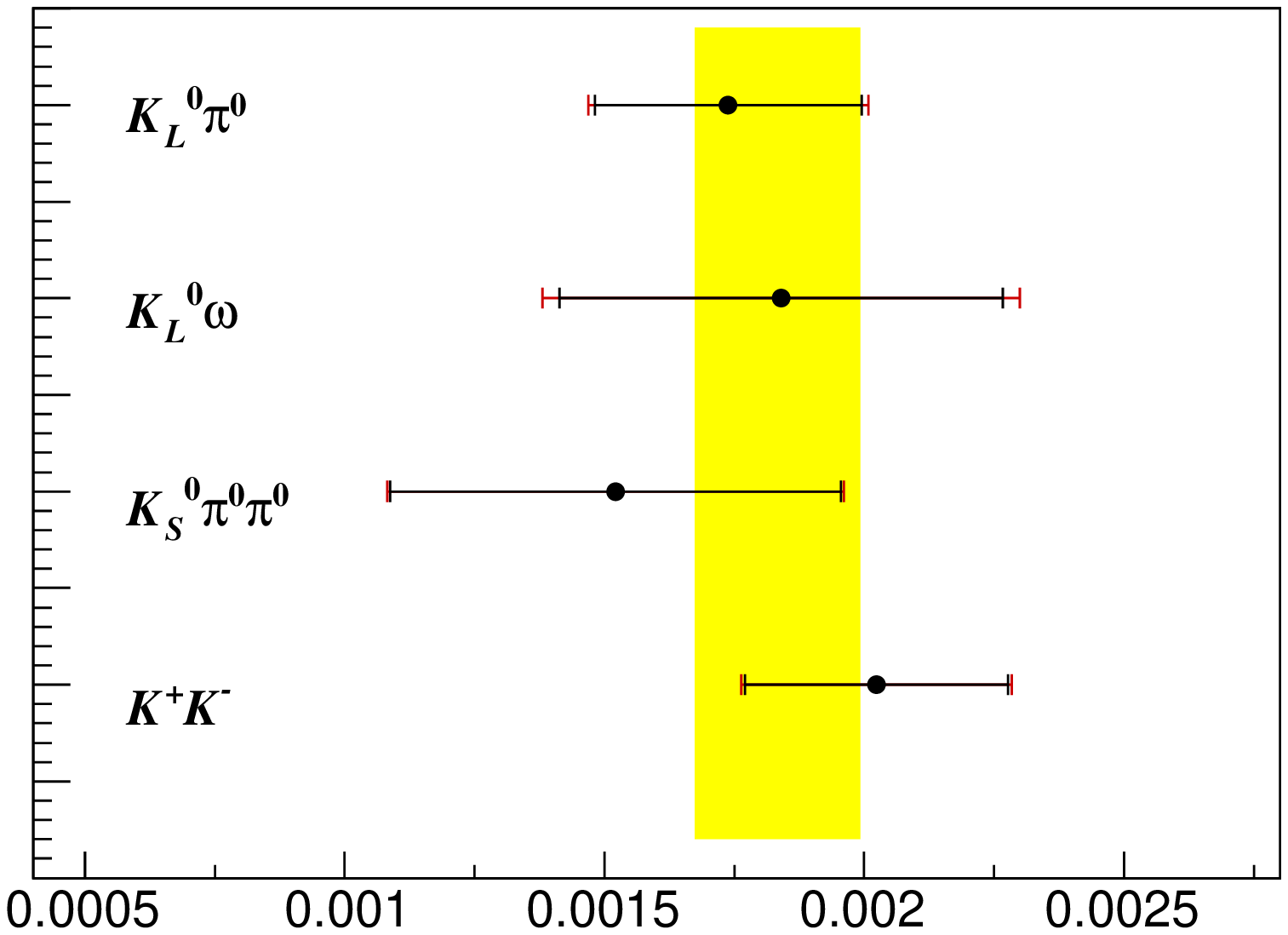}\put(75,55){(b)}\put(49,0){$N^-$}\end{overpic}
\caption[]{
    The results for $N^+$ (a) and $N^-$ (b).
	The wide red error bars show the total uncertainties, and the inner black error bars the statistical uncertainties. The yellow bands correspond to the one-$\sigma$ regions around the mean values.
}
\label{fig:CPtagFit}
\end{center}
\end{figure}

The $N^+$ and $N^-$ parameters are determined for each tag channel, and systematic uncertainties are assigned as discussed in Sec.~\ref{sec:systematics}.  The results are displayed in  Fig.~\ref{fig:CPtagFit}.
A least-squares fit, which takes account of the correlations between the systematic uncertainties, returns
\begin{eqnarray}
\langle N^+ \rangle &=& (4.73 \pm 0.20 \pm 0.09) \times 10^{-3} \, , \nonumber \\
\langle N^- \rangle &=& (1.83 \pm 0.16 \pm 0.04) \times 10^{-3} \, , \nonumber 
\end{eqnarray}
where, here and for all subsequent results, the first uncertainty is statistical and the second is systematic.  The $\chi^2$ per number of degrees of freedom for the $N_+$ and $N_-$ fits are 6.68/6 and 1.21/3, respectively.
It follows from Eq.~(\ref{eq:cpDTST}) that 
\begin{equation}
\cpfpi = 0.721 \pm 0.019 \pm 0.007 \, , \nonumber 
\end{equation}
indicating that the decay $D\to\fpi$ is predominantly $C\!P$-even.

\subsection{$D \to \pipi\piz$ tag}

A second determination of $\cpfpi$ is made using the quasi $C\!P$-even eigenstate $D \to \pipi\piz$.   From the ST and DT yields it follows that
\begin{equation}
    N^{\pi\pi\piz} = (1.64 \pm 0.21 \pm 0.06) \times 10^{-3}  \, , \nonumber 
\end{equation}
which, as expected, lies very close to the result for  $\langle N^- \rangle$.
\noindent Using Eq.~(\ref{eq:pipipizDTST}) and the measured value of $\cpf^{\pi\pi\piz}$~\cite{Malde:2015mha} it is found that
\begin{equation}
\cpfpi = 0.753 \pm 0.028 \pm 0.010 \, , \nonumber
\end{equation}
\noindent together with a correlation coefficient of 0.15 with the measurement performed with the $C\!P$-eigenstate tags, mainly arising from the common use of $\langle N^+ \rangle$ in the two determinations.

\subsection{$D \to \ksl\pipi$ tags}

The DT events tagged with $D \to \ksl\pipi$ decays are analyzed in phase-space bins of the tag channel.  The Dalitz plots of these decays have axes corresponding to the squared invariant masses $m^2_- = m(\ksl \pi^-)^2$ and $m^2_+ = m(\ksl \pi^+)^2$. Eight pairs of bins are defined symmetrically about the line $m_-^2 = m_+^2$ with a positive bin number in the region $m_+^2 > m_-^2$ and a negative one on the other side of the the line of equality.  
Bin boundaries have been defined according to the ``equal $\Delta \delta_{D}$ scheme'' of Ref.~\cite{CLEO:2010iul}, such that the strong-phase difference between symmetric points in the Dalitz plot spans an equal range in each bin, according to the expectations of an amplitude model~\cite{Aubert:2008bd}.  
For $D \to \ks \pipi$, both the CLEO and BESIII collaborations have made measurements of $c_i$, the cosine of the average strong-phase difference weighted by the $\Dz$ decay amplitude in each bin, and $c_i^\prime$, the analogous quantity for $D \to \kl \pipi$ decays~\cite{CLEO:2010iul,BESIII:2020lpk,BESIII:2020hlg}.  The current analysis uses the averaged BESIII and CLEO results for $c_i$ and $c_i^\prime$, together with the BESIII results for $K_i$ and $K_i^{\prime}$, which are the probabilities of the $\Dz$ decay occurring in bin $i$.  All of these inputs are reported in Ref.~\cite{BESIII:2020lpk}.

As is clear from Eq.~(\ref{eq:KsKlDT}), the expected yields $M_i^{(\prime)}$ are  symmetric under the exchange $i\leftrightarrow -i$ and so the yields in these pairs of bins are aggregated in the analysis, giving eight effective bins, labeled $1$ through to $8$, for each tag mode.
The DT yields are determined from independent fits to the events in each effective bin following the procedure already described  in  Sec.~\ref{sec:evtsel}.  In some regions, the sample sizes are low and it is necessary to fix the parameters of the convolving Gaussian function
to those determined from a fit to the whole of phase space.  The fitted yields include contamination from peaking background, which is estimated from MC simulation to be at the level of 2.3\% and 9.5\% for $D \to \ks\pipi$ and $D \to \kl\pipi$, respectively.

To improve the resolution of the location of the decay in the Dalitz plot, and thereby ensure the most reliable assignment of the phase-space bin, the tag decay is refitted with the $D$ mass constrained to the known mass of the $\Dz$ meson~\cite{pdg}.  Even with this constraint in place, there are occasional bin misassignments. These misassignments, and the variations in relative efficiency between bins, are accounted for by an $8 \times 8$ efficiency matrix  $\epsilon_{ij}$, which is determined from MC simulation: 
\beq
\epsilon_{ij}=\frac{N^{\rm rec}_{ij}}{N^{\rm gen}_i},
\edq
where $N^{\rm rec}_{ij}$ is the number of signal MC events generated in the $i$-th bin and reconstructed in the $j$-th bin, and $N^{\rm gen}_i$ is the number of the signal MC events generated in the $i$-th bin. The probability of misassignment varies from 5\% to 20\% depending on the bin and tag channel. The relative variation in efficiency between bins is within 20\%. Separate efficiency matrices $\epsilon^{\ks \pi\pi}_{ij}$ and $\epsilon^{\kl \pi\pi}_{ij}$ are determined for each tag mode. The full efficiency matrices can be found in the Appendix. 

\begin{figure}[!htp]
\begin{center}
\begin{overpic}[width=0.45\textwidth]{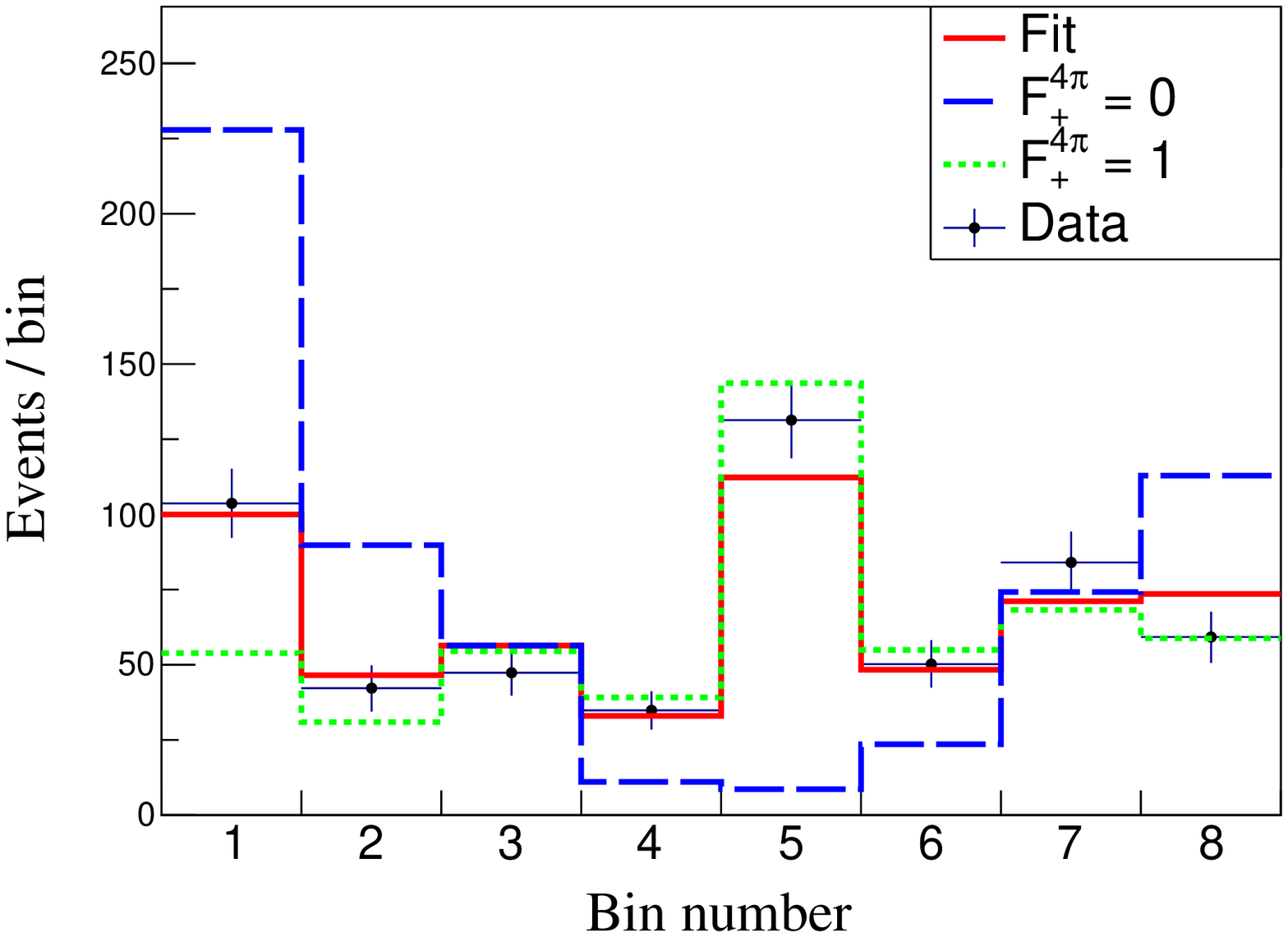}\put(40,55){(a)}\end{overpic}
\begin{overpic}[width=0.45\textwidth]{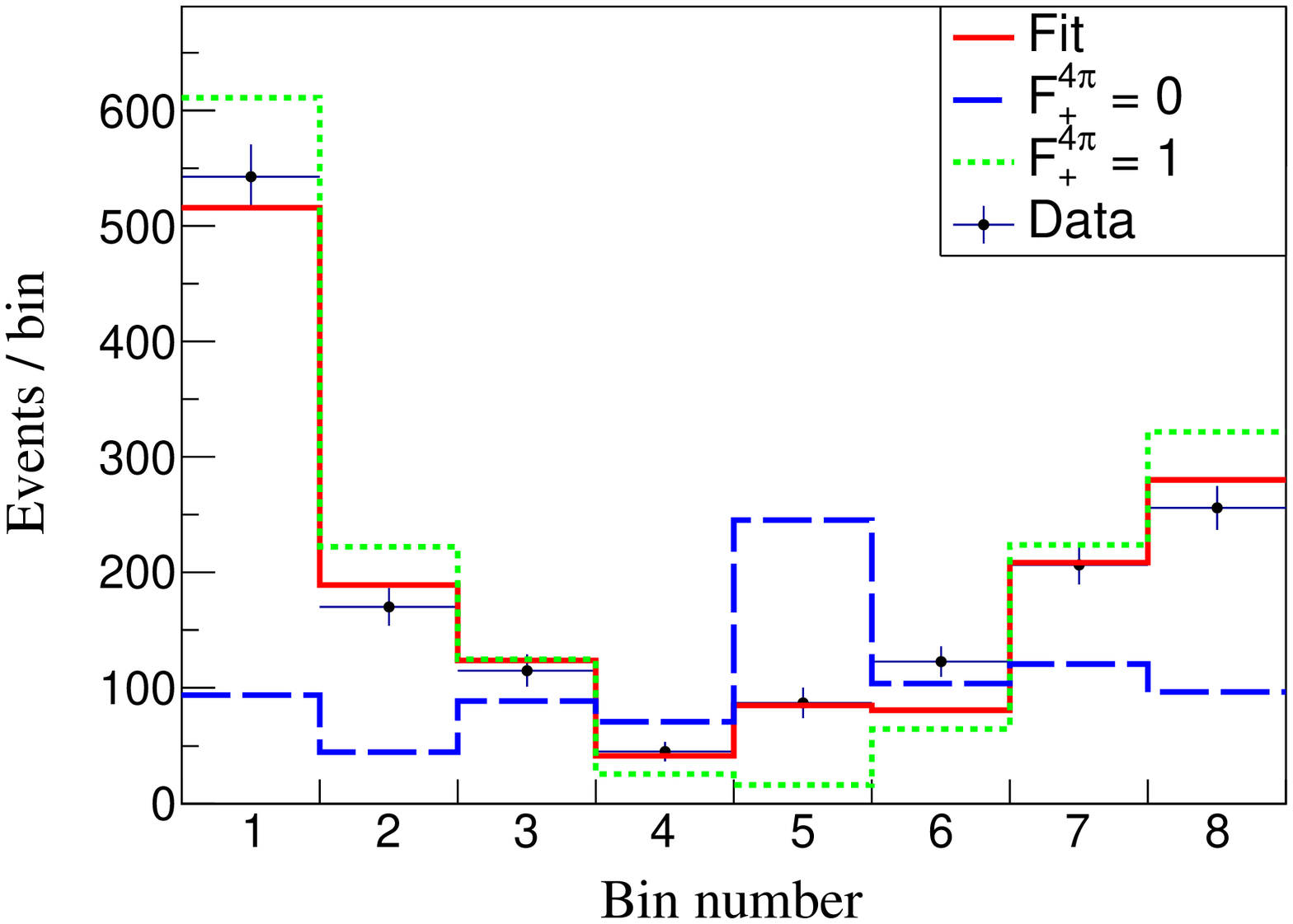}\put(40,55){(b)}\end{overpic}
\caption[]{
	Fit results for $D \to \ks\pipi$ (a) and $D \to \kl\pipi$ (b) channels.  Also shown are the expected distributions in the limiting  cases $\cpfpi = 0$ and $\cpfpi= 1$.
}
\label{fig:KsKlFit}
\end{center}
\end{figure}

The efficiency matrices are used to 
adjust the idealized expressions of Eq.~(\ref{eq:KsKlDT}), giving  expected DT signal yields in bin $i$ of 
\begin{eqnarray}
\label{eq:KsKlDTReal}
M_i(\Fpi,\ks\pipi)  = \hspace*{4.5cm} & \nonumber \\
H\sum_{j}^{8} \epsilon_{ij}^{\ks\pi\pi}[K_j+K_{-j} - 2\sqrt{K_jK_{-j}}c_j(2\cpfpi-1)] \, , \hspace*{0.1cm} &  \nonumber \\
M^\prime_i(\Fpi,\kl\pipi)  = \hspace*{4.5cm} & \nonumber \\
H^\prime\sum_{j}^{8} \epsilon_{ij}^{\kl\pi\pi}[K^\prime_j+K^\prime_{-j} + 2\sqrt{K^\prime_jK^\prime_{-j}}c^\prime_j(2\cpfpi-1)] \, . \hspace*{0.0cm} & \nonumber \\
\end{eqnarray}
The $C\!P$-even fraction is then  determined by minimizing the negative log-likelihood 
\beq
\label{eq:Likelihood}
\begin{split}
 -2\log\mathcal{L} =  
& -2 \log\,G(M^{\textrm{obs}}_i;M^{\textrm{exp}}_i,\sigma_{M_i})  \\
& -2 \log\,G(M^{\prime \, \textrm{obs}}_i;M^{\prime \, \textrm{exp}}_i,\sigma_{M^\prime_i}) \, , \\
\end{split}
\edq
where $G$ is a Gaussian function, 
$M_i^{(\prime) \, \rm obs}$ and $M_i^{(\prime) \, \rm exp}$ are the observed and predicted DT yields in bin $i$, respectively (here including peaking background contributions), and 
$\sigma_{M^{(\prime)}_i}$ is the uncertainty on  $M^{(\prime)\,\textrm{obs}}_i$. This fit yields 
\begin{equation}
\cpfpi = 0.754 \pm 0.031 \pm 0.009 \, . \nonumber 
\end{equation}
Separate fits to the $D \to \ks\pipi$ and $D \to \kl\pipi$ samples give the results  $0.813 \pm 0.045$ and $0.712 \pm 0.038$, respectively, with purely statistical uncertainties.  These fits are shown in Fig.~\ref{fig:KsKlFit}. 

\subsection{Assignment of systematic uncertainties}
\label{sec:systematics}

There are several sources of potential systematic bias on the three determinations of $\cpfpi$, some of which are correlated between two or all of the measurements.  The assigned uncertainties are summarized in Table~\ref{tab:CPFSys} and are discussed below.

\begin{table}[!hbp]
\renewcommand\arraystretch{1.5}
\caption{Summary of systematic contributions for the $\cpfpi$ measurement performed with $C\!P$ eigenstate, $D \to \pipi\piz$ and $D \to \ksl\pipi$ tags. The entry `/' indicates that there is no possible bias from the listed source. \vspace*{0.2cm}}
\label{tab:CPFSys}
\centering
\begin{tabular}{  l c c c  }
  \toprule
  Source      & \hspace*{0.5cm}$C\!P$\hspace*{0.5cm}  & $\pipi\piz$ & $\ksl\pipi$  \\  
  \colrule
  ST and DT yields & 0.005 & 0.003 & / \\  
  Efficiency factorization  & 0.003 & 0.006 &  / \\
  $\cpf^{\pi\pi\piz}$ & / & 0.007 & / \\
  $K_i^{(\prime)}$ & / & / & 0.005 \\
  $c_i^{(\prime)}$ & / & / & 0.004 \\
  Bin misassignment & / & / & 0.003 \\
  $\ksl\pipi$ backgrounds & / & / & 0.003 \\
    $\ks$ veto & 0.003 & 0.003 & 0.003  \\
  \colrule
  Total & 0.007 & 0.010 & 0.009 \\
  \colrule

  \botrule
\end{tabular}
\end{table}

There are systematic uncertainties on $N_{\rm ST}$ and $N_{\rm DT}$ for each tag mode that lead to corresponding uncertainties in the values of $\langle N^+\rangle$, $\langle N^- \rangle$, and $N^{\pi\pi\pi^0}$.
There are various components contributing to these uncertainties.
To assess any bias associated with the fit procedure, alternative fits are performed to the $m_{\rm BC}$ distributions for the ST samples and fully reconstructed DT samples,  with variations of $\pm 0.5$~MeV applied to the end point of the ARGUS function.  The contributions of the peaking backgrounds in all samples are varied according to the uncertainties of  their branching fractions~\cite{pdg}.  The variation in results is taken as  the uncertainty associated with these sources. The effective ST yields for the $D \to \kl X$ tags have a small uncertainty associated with the total size of the charm-meson sample~\cite{BESIII:2018iev}.  The uncertainties arising from the knowledge of the $D \to \kl X$ branching fractions are also small, as those contributions to these uncertainties associated with the reconstruction efficiencies~\cite{BAM:567} are common to the $\epsilon'_{\,\textrm{ST}}(\kl X)$ efficiencies in Eq.~\ref{eq:KlST}, and so cancel in the measurement of the effective ST yields.   The  determinations of $N^{\rm ST}$ and $N^{\rm DT}$ are the dominant source of systematic  uncertainty for the $C\!P$ eigenstate tags.  The common use of $\langle N^- \rangle$ means that this uncertainty is correlated between the $C\!P$-eigenstate measurement and that made with $D \to \pi^+\pi^-\pi^0$ tags.

The measurements performed with the $C\!P$-eigenstate and the $D \to \pi^+\pi^-\pi^0$ tags assume that the DT-event efficiency can be factorized into the reconstruction efficiencies of the two decay modes in a universal manner for all tags.  Studies with MC simulation show that this is not exactly true, and so small correction factors obtained from the simulation are applied to each class of DT in the measurement.  Shifts of $0.06 \times 10^{-3}$ and $0.01 \times 10^{-3}$ are observed in the values of $\langle N^+\rangle$ and $\langle N^-\rangle$, respectively, and of $0.05 \times 10^{-3}$ for $N^{\pi\pi\piz}$, which are assigned as systematic uncertainties that when propagated to $\cpfpi$ give the uncertainties listed in Table~\ref{tab:CPFSys}.

The quasi-$C\!P$-eigenstate measurement carries an uncertainty from the imperfect knowledge of the $C\!P$-even fraction of $D \to \pipi\piz$~\cite{Malde:2015mha}.  This is the dominant systematic uncertainty for the determination of $\cpfpi$ using this tag channel.

The measurement performed with $D \to \ksl\pipi$ tags has uncertainties associated with the knowledge of the external parameters $K_i^{(\prime)}$ and $c_i^{(\prime)}$.  To evaluate the size of these contributions, the measurement is re-performed 1000 times for each set of inputs with the values of the parameters smeared in a manner that corresponds to the covariance matrices reported in Ref.~\cite{BESIII:2020lpk}.    The spread on the resulting distribution for $\cpfpi$ over the ensemble of measurements is assigned as the uncertainty of the result from these sources.   
The efficiency matrix, used to describe bin misassignment and relative bin-to-bin efficiencies, has uncertainties arising from the finite size of the MC sample from which its elements are determined.  The effect of these uncertainties, which are most important for bin misassignments, are determined by re-performing the fit 1000 times with smeared inputs. 
The branching fractions of the peaking backgrounds in these samples are varied within their measured uncertainties to determine the size of the potential bias from this source~\cite{pdg}.

A final uncertainty is assigned, common to all three measurements, to account for the possibility that the $\ks$ veto in the DT selection leads to a sample with a different $C\!P$-even fraction to that of the unbiased decay.  The size of the effect is  estimated to be $3 \times 10^{-3}$ and is determined from the amplitude model by computing $\cpfpi$ with and without the relevant region of phase space included.  
 
\subsection{Combination of results}
\label{sec:combination}

The results of the three separate measurements of $\cpfpi$ are summarized in Table~\ref{tab:CPFall}.  It can be seen that these results are consistent with each other.   A combination is made of these measurements, 
taking account of the correlation coefficient of 0.15 between the $C\!P$-eigenstate and $D\to\pipi\piz$ tags, 
that yields the result presented in Table~\ref{tab:CPFall}.  
This result is consistent with the CLEO-c result of $0.737 \pm 0.028$ from a measurement that employs a similar analysis strategy and is around twice as precise~\cite{Malde:2015mha}. 
It is also compatible with a CLEO-c determination that is based on localized measurements in phase space and has around a 30\% smaller uncertainty~\cite{Harnew:2017tlp}.

\begin{table}[!hbp]
\renewcommand\arraystretch{1.5}
\caption{Summary of $\cpfpi$ results from the different tag channels and the combination of these results, where the first uncertainty is statistical and the second is systematic.\vspace*{0.2cm}}
\label{tab:CPFall}
\centering
\begin{tabular}{  l c   }   
  \toprule
  Tag modes      & $\cpfpi$  \\  
  \colrule
  $C\!P$ eigenstates   &  0.721$\pm$0.019$\pm$0.007 \\
  $D \to \pipi\piz$      &  0.753$\pm$0.028$\pm$0.010 \\  
  $D \to \ksl\pipi$ &  0.754$\pm$0.031$\pm$0.009 \\  
  \colrule                                                       
  Combination            &  0.735$\pm$0.015$\pm$0.005 \\
  \botrule
\end{tabular}
\end{table}

\section{SUMMARY}

In summary, a measurement has been made of $\cpfpi$, the $C\!P$-even fraction of the decay $\Dz\to\fpi$, using 2.93~$\invfb$ of $e^+e^- \to \psipp \to D\bar{D}$ data  collected by the BESIII experiment. The measurement is the combination of three separate, but consistent, determinations of this parameter made with $C\!P$-eigenstate tags,  the quasi-$C\!P$-eigenstate tag $D \to \pipi\piz$ and a study of the distributions in phase space of $D \to \ksl\pipi$ tags. The $C\!P$-even fraction is determined to be  
\begin{equation}
\cpfpi = 0.735 \pm 0.015 \pm 0.005 \, , \nonumber
\end{equation}
where the first uncertainty is statistical and the second is systematic. This result is compatible with previous measurements~\cite{Malde:2015mha,Harnew:2017tlp}, but has a higher precision, and provides valuable input to measurements of the CKM angle $\gamma$ and charm-mixing studies performed at LHCb and Belle~II.

The data set can be further exploited to study the $C\!P$ content and related parameters in localized regions of phase space, as was demonstrated in Ref.~\cite{Harnew:2017tlp}.  Such measurements, and improved determinations of $\cpfpi$, will be particularly interesting with the substantially larger samples that will be collected at BESIII over the coming years~\cite{BESIII:2020nme}.

\section*{ACKNOWLEDGMENTS}
The BESIII collaboration thanks the staff of BEPCII and the IHEP computing center and the supercomputing center of USTC for their strong support. This work is supported in part by National Key R\&D Program of China under Contracts Nos. 2020YFA0406400, 2020YFA0406300; National Natural Science Foundation of China (NSFC) under Contracts Nos. 11335008, 11625523, 11635010, 11735014, 11835012, 11935015, 11935016, 11935018, 11961141012, 12022510, 12025502, 12035009, 12035013, 12192260, 12192261, 12192262, 12192263, 12192264, 12192265, 11705192, 11950410506, 12061131003, 12105276, 12122509; the Chinese Academy of Sciences (CAS) Large-Scale Scientific Facility Program; Joint Large-Scale Scientific Facility Funds of the NSFC and CAS under Contract No. U1732263, U1832207, U1832103, U2032111; CAS Key Research Program of Frontier Sciences under Contract No. QYZDJ-SSW-SLH040; 100 Talents Program of CAS; INPAC and Shanghai Key Laboratory for Particle Physics and Cosmology; ERC under Contract No. 758462; European Union's Horizon 2020 research and innovation programme under Marie Sklodowska-Curie grant agreement under Contract No. 894790; German Research Foundation DFG under Contracts Nos. 443159800, Collaborative Research Center CRC 1044, GRK 2149; Istituto Nazionale di Fisica Nucleare, Italy; Ministry of Development of Turkey under Contract No. DPT2006K-120470; National Science and Technology fund; National Science Research and Innovation Fund (NSRF) via the Program Management Unit for Human Resources \& Institutional Development, Research and Innovation under Contract No. B16F640076; STFC (United Kingdom); Suranaree University of Technology (SUT), Thailand Science Research and Innovation (TSRI), and National Science Research and Innovation Fund (NSRF) under Contract No. 160355; The Royal Society, UK under Contracts Nos. DH140054, DH160214; The Swedish Research Council; U. S. Department of Energy under Contract No. DE-FG02-05ER41374.

\section*{APPENDIX: EFFICIENCY MATRICES}

The efficiency matrices for $\fpi$ vs. $\ksl\pipi$ giving the efficiency in each bin and migration probabilities between bins are shown in Table~\ref{tab:CPFEffM}.

\begin{table*}[!hbp]
\caption{Efficiency matrix $\epsilon_{ij}$ (\%) for $\fpi$ vs. $\ksl\pipi$. The column gives the true bin $i$, while the row gives the reconstructed bin $j$.}
\label{tab:CPFEffM}
\centering
\begin{tabular}{ c c c c c c c c c}
	\hline
	Bins   & 1 & 2 & 3 & 4 & 5 & 6 & 7 & 8  \\
	\multicolumn{9}{c}{$\epsilon_{ij}$ for $\fpi$ vs. $\ks\pipi$} \\
	\hline
  1 & 0.2928 & 0.0103 & 0.0014 & 0.0005 & 0.0012 & 0.0009 & 0.0013 & 0.0139 \\  
  2 & 0.0219 & 0.2965 & 0.0121 & 0.0003 & 0.0012 & 0.0006 & 0.0015 & 0.0026 \\  
  3 & 0.0029 & 0.0089 & 0.3424 & 0.0062 & 0.0023 & 0.0010 & 0.0008 & 0.0009 \\  
  4 & 0.0015 & 0.0005 & 0.0110 & 0.3394 & 0.0080 & 0.0015 & 0.0008 & 0.0011 \\  
  5 & 0.0020 & 0.0004 & 0.0011 & 0.0035 & 0.3070 & 0.0074 & 0.0010 & 0.0013 \\  
  6 & 0.0023 & 0.0005 & 0.0013 & 0.0008 & 0.0156 & 0.2810 & 0.0176 & 0.0020 \\  
  7 & 0.0028 & 0.0010 & 0.0009 & 0.0006 & 0.0016 & 0.0134 & 0.2698 & 0.0186 \\  
  8 & 0.0252 & 0.0019 & 0.0010 & 0.0004 & 0.0010 & 0.0012 & 0.0188 & 0.2701 \\  
	\hline
	\multicolumn{9}{c}{$\epsilon_{ij}$ for $\fpi$ vs. $\kl\pipi$} \\
  1 & 0.4216 & 0.0167 & 0.0010 & 0.0002 & 0.0005 & 0.0008 & 0.0028 & 0.0276 \\  
  2 & 0.0399 & 0.4139 & 0.0150 & 0.0005 & 0.0008 & 0.0012 & 0.0024 & 0.0046 \\  
  3 & 0.0026 & 0.0177 & 0.4366 & 0.0073 & 0.0018 & 0.0009 & 0.0030 & 0.0023 \\  
  4 & 0.0015 & 0.0026 & 0.0196 & 0.4497 & 0.0107 & 0.0026 & 0.0030 & 0.0007 \\  
  5 & 0.0002 & 0.0003 & 0.0009 & 0.0060 & 0.4164 & 0.0150 & 0.0016 & 0.0009 \\  
  6 & 0.0039 & 0.0015 & 0.0011 & 0.0013 & 0.0174 & 0.3653 & 0.0371 & 0.0045 \\  
  7 & 0.0071 & 0.0016 & 0.0006 & 0.0004 & 0.0010 & 0.0164 & 0.3753 & 0.0415 \\  
  8 & 0.0476 & 0.0039 & 0.0006 & 0.0001 & 0.0005 & 0.0013 & 0.0364 & 0.3683 \\  
	\hline
\end{tabular}
\end{table*}

\bibliographystyle{BESIII}
\bibliography{references}


\end{document}